\documentclass[3p,times,twocolumn]{elsarticle}

\usepackage{ecrc}

\volume{00}

\firstpage{1}

\journalname{Nuclear Physics B Proceedings Supplement}

\runauth{P. Mitra}

\jid{nuphbp}

\jnltitlelogo{Nuclear Physics B Proceedings Supplement}

\usepackage{amssymb}

\usepackage[figuresright]{rotating}

\newcommand{\bb}{\begin{eqnarray}}
\newcommand{\ee}{\end{eqnarray}}

\begin{document}

\begin{frontmatter}



\dochead{}

\title{
BLACK HOLE ENTROPY WITH AND WITHOUT LOG CORRECTION IN LOOP QUANTUM GRAVITY}


\author{P. Mitra}

\address{{Saha Institute of Nuclear Physics, Calcutta 700064}} 

\begin{abstract}

Earlier calculations of black hole entropy in loop quantum gravity have given a
term proportional to the area  with a correction
involving the logarithm of the area when
the area eigenvalue is close to the classical area.
However the calculations yield an entropy proportional to the area
eigenvalue with no such correction when the area eigenvalue is 
large compared to the classical area.
\end{abstract}


\end{frontmatter}


\section*{Introduction}

Black holes were found to satisfy an area law,
with the area tending to increase quite generally.
This was reminiscent of entropy, and since black hole
horizons imply limitations on information, it was
argued that the area could indeed be a measure of
some black hole entropy.

Scalar field theory in a
black hole background was then found to lead to an
emission of particles at a temperature
\bb
T={\hbar\kappa\over 2\pi }
\ee
related to the surface gravity $\kappa$.
This established a precise connection  of  black  holes
with  thermodynamics.

Consider the grand partition function  
for euclidean charged black holes:
\bb
Z_{\rm grand}\equiv e^{-{M- TS-\phi Q\over T}}\approx e^{-I/\hbar}.
\ee
The functional  integral over  all   configurations  
consistent with appropriate boundary conditions is 
semiclassically approximated  by  the integrand, which
involves the action $I$. For a
euclidean Reissner - Nordstr\"{o}m  black  hole,
\bb
I={1\over 2}\beta(M-Q\phi)={A\over 4}.
\ee
Hence
\bb
M=T(S+{I\over \hbar})+\phi Q=T(S+{A\over 4\hbar})+\phi Q.
\ee
There is a formula called the Smarr formula:
\bb
M={\kappa A\over 4\pi}+\phi Q
=T{A\over 2\hbar}+\phi Q.
\ee
It implies that $S={A\over 4\hbar}$.


{For the extremal black hole}, $r_+=r_-, Q=M, \phi=1$
This is of special interest:
the topology changes discontinuously in the passage from
a (euclidean) non-extremal to an extremal black hole.
The action is
\bb
I={1\over 2}\beta(M-Q\phi)=0,
\ee
\bb
M=T(S+{I\over \hbar})+\phi Q=TS+M\Rightarrow S=0
\ee
Here, the quantum theory is based exclusively on
the extremal topology.

{An alternative quantization} involves a {\it sum over topologies}.
Here a temperature $\beta^{-1}$ and a chemical
potential $\phi$ are specified as inputs at the boundary of manifold, and 
the mass $M$ and the charge $Q$ of the black hole are calculated
as functional integral averages.
The definition of extremality $Q=M$ is imposed on these averages. This is
{\it extremalization after
quantization}, as opposed to {\it quantization after
extremalization.}
The partition function is of the form
\bb
\sum_{\rm topologies}\int d\mu e^{-I},
\ee
with   $I$   appropriate for
non-extremal/extremal topology.
The semiclassical approximation involves
replacing the integral by the
maximum value of the integrand.
That occurs for the {non-extremal} case,
so that once again
$S={A\over 4\hbar}$.

\section*{Counting states in loop quantum gravity}

A framework for the calculation of black hole entropy
is provided by loop
quantum gravity or the quantum geometry approach. 

Quantum states are
built up by associating spin variables with ``punctures'' on 
a horizon.
The entropy is obtained by counting all possible configurations of punctures 
consistent with a given horizon area {\it i.e.,} a particular
eigenvalue of the area {\it operator}).

A generic configuration has $s_j$ punctures with spin $j,j=1/2,1,3/2....$
\bb 2\sum_js_j\sqrt{j(j+1)}=A,\ee 
$A$ being the eigenvalue of the horizon area operator in units with 
\bb 4\pi\gamma\ell_P^2=1,\ee 
$\gamma$ being the Barbero-Immirzi parameter, $\ell_P$ the Planck length.
There is a {\it spin projection constraint}
\bb
\sum m=0 {\rm ~mod~}\frac{k}{2},\quad {\rm all~punctures},
\ee
\bb
m\in\{-j, -j+1,... j\} {\rm ~for~puncture~with~spin~}j
\ee
Here $k$ is an integer representing the level of the Chern-Simons theory,
equal to the {classical horizon area} in the units defined above..

For simplicity, we first consider spin 1/2 on {\it each} puncture.
The punctures have to be considered as distinguishable.
If the number of punctures is $p$, all with spin 1/2,
\bb
2p\sqrt{\frac34}=A,
\ee
so {\it if we neglect the spin projection constraint}, the entropy is
\bb
\ln 2^p=p\ln 2={A\ln 2\over \sqrt 3}=
{A\ln 2\over 4\sqrt 3\pi\gamma\ell_P^2}.
\ee
This involves $\gamma$, which can be {\it chosen}
to yield the desired Bekenstein-Hawking entropy
${A\over 4\ell_P^2}$:
\bb
\gamma={\ln 2\over \sqrt 3\pi}.
\ee
This assumes only $j=1/2$ at each puncture.
For a general configuration, there may be $s_j$ punctures with spin $j$ 
with different $j$.
\bb N= {(\sum_j s_j)!\over \prod_j
s_j!}\prod_j (2j+1)^{s_j}\ee 
if the $\sum m$ constraint is neglected.
Here the first factor is the number of ways of choosing
the locations of spins, the second factor counts
the number of spin states at different punctures.
One must sum $N$ over all nonnegative
$s_j$ consistent with a given $A$.
To estimate the sum by maximizing $\ln N$,
one has to vary $s_j$ at fixed $A$.
The simplified Stirling approximation {$\ln p!\simeq p\ln p -p$}
yields 
\bb
\ln N=\sum_j s_j\ln
{2j+1\over s_j} + (\sum_j s_j)\ln (\sum_j s_j),
\ee
\bb
\delta\ln N=\sum_j\delta s_j\Big[\ln (2j+1)-\ln
s_j+\ln\sum_ks_k\Big].
\ee
With some Lagrange multiplier
$\lambda$ to enforce the area constraint, 
\bb\ln
(2j+1)-\ln s_j+\ln\sum_ks_k=\lambda\sqrt{j(j+1)},\ee 
\bb
s_j=(2j+1)\exp\Big[-\lambda\sqrt{j(j+1)}\Big]\sum_ks_k\;.\ee 
Summing over $j$ yields
\bb\sum_j(2j+1)\exp\Big [-\lambda\sqrt{j(j+1)}\Big]=1,\ee 
which determines
$\lambda\approx$  1.72 and 
\bb\ln N=\lambda A/2.\ee 
To make this ${A\over 4\ell_P^2}$ 
the Barbero-Immirzi parameter
\bb
\gamma=\lambda /(2\pi)\approx 0.274.
\ee
Summing over $s_j$ may raise this 
value, the projection constraint will lower it.

To enforce the neglected constraint $\sum m=0$,
note that if $p$ is odd, there is no such state,
but if $p$ is even, the number of such
states is ${}^pC_{p/2}$.
For large $p$, one may use the {\it full} Stirling approximation: 
$\ln p!\simeq  p\ln p -p +\frac{1}{2}\ln (2\pi p)$,
\bb
\ln {p!\over (p/2)!(p/2)!}\simeq p\ln 2-\frac{1}{2}\ln p,
\ee
\bb
S\simeq {A\over 4\ell_P^2}-\frac12\ln A.
\ee
The spin projection constraint contains a mod $\frac12 k$,
which has been ignored so far. 
But it becomes relevant for large eigenvalues
of the area operator at a fixed classical area.
In the pure spin 1/2 case, with an
even number $p$ of such spins,
the total number of spin states is
\bb
2^p=\sum_{r=0}^p{}^pC_r.
\ee
Let $p>k=2n$ and for simplicity, let $k$ be even. 
Define $\ell$ by
\bb
\frac{p}{2}=\ell ~{\rm mod}~ n,\quad 0\leq\ell\leq n-1.
\ee
If $\ell$ spins are up and $p-\ell$ spins are down, 
$$\sum m=\ell-\frac{p}{2}$$
which is zero mod $n$, as required by spin projection condition.
There are also other possibilities: $\ell +n, \ell+2n,...,p-\ell$. 

The total number of ways for spin projection zero modulo $k/2=n$ is 
\bb
N={}^pC_{\ell}+{}^pC_{\ell+n}+...+{}^pC_{p-\ell-n}+{}^pC_{p-\ell}.
\ee
Now, for  $0<s<n-1$, one has the identity
\bb
(1+e^{2is\pi/n})^p=\sum_{r=0}^pe^{2irs\pi/n}~{}^pC_r.
\ee
Hence,
\bb
e^{-2i\ell
s\pi/n}(1+e^{2is\pi/n})^p=\sum_{r=0}^pe^{2i(r-\ell)s\pi/n}\;{}^pC_r.
\ee
We have to add these equations for all values of $s$. 

Only those coefficients of ${}^pC_r$ survive which have $r=\ell$ 
modulo $n$:
\bb
\sum_{s=0}^{n-1}e^{-2i\ell\pi s/n}(1+e^{2is\pi/n})^p=nN.
\ee
For fixed $n$ and large $p$, the sum is
dominated by the term of highest magnitude. But
\bb
1+e^{2is\pi/n}=2\cos (s\pi/n)e^{is\pi/n}.
\ee
The highest magnitude occurs for $s=0$ and
\bb
N\approx 2^p/n,
\ee
other terms being exponentially suppressed for large $p$ at fixed $n$.
Now the area eigenvalue is
\bb
4\pi\gamma\ell_P^2\sqrt{3}p\equiv A, 
\ee
and
\bb
S=\log N= A{\log 2\over 4\pi\gamma\ell_P^2\sqrt{3}}-\log n.
\ee
For fixed $n\sim$ classical horizon area, this goes like $A$.

In earlier calculations, $n\sim p$ and this argument does not hold.

Now we come to the case of arbitrary spins.
Let $s_{jm}$ be the number of punctures with spin quantum numbers $j,m$
in a certain configuration.
The no. of all spin states is
\bb
\sum_{\{ s_{jm}\} }{(\sum_{jm} s_{jm})!\over \prod_{jm} (s_{jm}!)}.
\ee
Not all are allowed by the spin projection condition
\bb
\sum_{jm}ms_{jm}=0,
\ee
where strict equality will be imposed at first,
other possibilities modulo $n$ being taken into account later.
States with definite area eigenvalue $A$ have
\bb
\sum_{jm}8\pi\gamma\ell_P^2\sqrt{j(j+1)}s_{jm}=A.
\ee
To maximize the probability of a configuration $\{s\}$, one must maximize
the combinatorial factor for $\{s\}$ or its logarithm:
\bb
(\sum \delta s)\ln\sum s -\sum (\delta s\ln s)=0,
\ee
where the simplified version of Stirling's approximation
{\it i.e., without} the square root factor is used. 
This relation is subject to
\bb
\sum m \delta s=0, \quad \sum\sqrt{j(j+1)}\delta s=0.
\ee
With two Lagrange multipliers, one finds
\bb
{s_{jm}\over \sum s}=e^{-\lambda\sqrt{j(j+1)}-\alpha m}.
\ee
It follows that
\bb
1=\sum_{jm}e^{-\lambda\sqrt{j(j+1)}-\alpha m}.
\ee
Up to this point calculations are independent of the total spin
projection.
The condition of vanishing spin projection sum implies that
\bb
\alpha=0.
\ee
Later we shall need a non-vanishing value of $\alpha$. 
The combinatorial factor for $\{ s\}$ reduces to
\bb
\exp({\lambda A\over 8\pi\gamma\ell_P^2})
\ee
in the simplified Stirling approximation, and there is an extra factor
\bb
{\sqrt{2\pi\sum\bar s}\over\prod\sqrt{2\pi\bar s}}
\ee
in the full Stirling approximation,
where ${\bar s_{jm}}$ is the most probable configuration.

To take care of this piece, it is necessary
to expand $s_{jm}$ about ${\bar s_{jm}}$
and sum over the fluctuations. 
Because of the stationary condition about most probable configuration, 
the first order variation vanishes and second order variations are kept:
\bb
{(\sum s)!\over \prod s!}={(\sum \bar s)!\over \prod \bar s!}
\exp[-\sum{(\delta s)^2\over 2\bar s}+ {(\delta\sum s)^2\over 2\sum\bar s}].
\ee
If the second term in the exponent were absent,
each $\delta s=s-\bar s$ would produce on integration a factor
$$\sqrt{2\pi\bar s},$$
to be compared to a similar factor in the denominator.
Note that the second term in the exponent produces a zero mode
given by $\delta s\propto s$,
but this is eliminated from the integration
because it is not consistent with the area constraint. 

Now there are two constraints on the $\delta s$,
so two factors are missing in the numerator. 
One has instead a factor $\sqrt{2\pi\sum\bar s}$ in the numerator.
It is easy to see that each $\bar s\propto A$, so 
each factor $\propto\sqrt A$, yielding a resultant
factor $\frac{1}{\sqrt A}$.

The number of states with spin projection zero is thus
\bb
N_0={1\over \sqrt{A}}\exp ({\lambda A\over 8\pi\gamma\ell_P^2}),
\ee
where constant factors have been ignored
and $\lambda$ is determined by the condition
\bb
\sum_{jm} e^{-\lambda\sqrt{j(j+1)}}=1,
\ee
given above.

To take into account the possibility of $\sum ms_{jm}$ being
equal to zero {\it modulo} $n$,
we let the spin projection be $M$, say. 

It is necessary to restore $\alpha\neq 0.$ The condition 
\bb
1=\sum_{jm}e^{-\lambda\sqrt{j(j+1)}-\alpha m}
\ee
cannot determine both parameters, but can be
solved in principle for $\lambda(\alpha)$. 
Note that $\bar s$ now depends
on $\alpha$ and the exponential factor in the number of configurations
changes to
\bb
\exp({\lambda(\alpha) A\over 8\pi\gamma\ell_P^2} +\alpha M)
\ee
The projection constraint now takes the form 
\bb
\sum me^{-\lambda\sqrt{j(j+1)}-\alpha m}=\frac{M}{\sum\bar s}.
\ee
So although $\alpha\neq 0$, it is small for $M\ll A$: 
\bb
\lambda'(0)=0,
\ee
\bb
{M\over A/(8\pi\gamma\ell_P^2)}=-\lambda'(\alpha)=
-\alpha\lambda''(0).
\ee
Furthermore,
\bb
{\lambda(\alpha) A\over 8\pi\gamma\ell_P^2} +\alpha M&=&
(\lambda(\alpha)-\alpha^2\lambda''(0)){A\over 8\pi\gamma\ell_P^2}\nonumber\\
&=& (\lambda(0)-{\alpha^2\over 2}\lambda''(0)) {A\over 8\pi\gamma\ell_P^2}\nonumber\\
&=&\lambda(0){A\over 8\pi\gamma\ell_P^2}-
\frac{M^2}{2\lambda''(0)}
{8\pi\gamma\ell_P^2\over A}\nonumber\\
\ee
Note that $\lambda(0)$ is what was called $\lambda$ earlier.
Since
\bb
\lambda''(0)=\frac{\sum m^2e^{-\lambda(0)\sqrt{j(j+1)}}}{\sum\sqrt{j(j+1)}e^{-\lambda(0)\sqrt{j(j+1)}}},
\ee
which is positive, independent of $A,M$ and $\sim o(1)$, 
we can write 
\bb
N_M= N_0e^{-4\pi\gamma\ell_P^2M^2/(\lambda''(0)A)}.
\ee
Since $M=0$ mod $n$, we have to sum $N_M$ over the values $rn$, where 
$r=0,\pm 1, \pm2,...,$ and there arises a factor
\bb
\sum_r e^{-4\pi\gamma\ell_P^2r^2n^2/(\lambda''(0)A)}
\ee
which, on approximation by an integral over $r$, 
is seen to involve a factor $\sqrt{A}/n$, 
cancelling the square root in $N_0$. 
Apart from a factor
$\frac{1}{n}$, which can be neglected, we find
\bb
N= \exp ({\lambda(0) A\over 8\pi\gamma\ell_P^2}),
\ee
implying that the entropy has no logarithmic correction.

In the earlier literature, the area eigenvalue was 
restricted to be close to the classical area and that
led to logarithmic correction terms. 
The only reason to consider eigenvalues close to classical area
was to check Bekenstein's proposal of classical
area as a measure of entropy. 
It worked out
up to logarithmic corrections.

The present calculations show that for area
eigenvalues much larger than the fixed classical area, 
the degeneracy is still exponential in the area eigenvalue
but it is a pure exponential and
the entropy ceases to have a logarithmic correction term.

\section*{Counting states: SU(2) theory}

In a non-standard SU(2) formulation of loop quantum gravity,
the number of states 
for a distribution of spins over punctures 
arises from properties of SU$_q$(2) as
\bb
{2\over k+2}{(\sum_j n_j)!\over\prod_j n_j!}
\sum_{a=1}^{k+1}\sin^2 {a\pi\over k+2}\prod_j 
\bigg[{\sin {a\pi(2j+1)\over k+2} \over\sin {a\pi\over k+2}}\bigg]^{n_j}.
\ee
If $j_i=1/2$ for each puncture, this is
\bb
N=\frac{2}{k+2}\sum_a\sin^2 {a\pi\over k+2} [2\cos {a\pi\over k+2}]^p.
\ee
If the level $k$ is large, 
one can approximate the sum by an integral:
\bb
N=\frac{2}{k+2}\int_0^k da \sin^2 {a\pi\over k}2^p\cos^p {a\pi\over k}.
\ee
For odd $p$, the integral vanishes.
For even $p$, 
\bb
N&=&\frac{4}{\pi}\int_0^{\pi/2} dx\ \sin^2 x\ 2^p\ \cos^p x\\
&=&\frac{4}{\pi}2^p\frac{\pi}{2}{(p-1)!!\over (p+2)!!}\\
&\approx &2^{p+3/2}p^{-3/2}\pi^{-1/2},\\
\log N&\approx& p\log 2-\frac32\log p.
\ee
Thus there is an area term
and a logarithm with coefficient -3/2.

Next we keep $k$ fixed when $p$ is made large.
The argument of the sine/cosine varies from term to term.
The finite sum is dominated by its largest term,
which occurs for largest values of 
$|\cos{a\pi\over k+2}|$.
The number of punctures $p$ occurs only in the exponent: 
\bb
N=\frac{4}{k+2}\sin^2 {\pi\over k+2} [2\cos {\pi\over k+2}]^p.
\ee
The area $\propto p$: 
\bb
\log N\propto {\rm area}. 
\ee

The argument can be extended, as in the U(1) case, to  general spins.
\bb
N&=&\frac{2}{k+2}{(\sum_j n_j)!\over\prod_j n_j!}
\sum_{a=1}^{k+1}\sin^2 {a\pi\over k+2}\prod_j [f_j]^{n_j},\nonumber\\
&=&\frac{2}{k+2}{(\sum_j n_j)!\over\prod_j n_j!}
\sum_a\sin^2 {a\pi\over k+2} F(\cos {a\pi\over k+2}),\nonumber\\
\ee
with 
\bb
f_j(\cos {a\pi\over k+2})\equiv
{\sin {a\pi(2j+1)\over k+2} \over\sin {a\pi\over k+2}}
\ee
and
\bb
F(\cos {a\pi\over k+2})\equiv\prod_j [f_j]^{n_j}.
\ee

Let us first consider $k$ becoming large,
so that the sum over $a$ can be treated as an integral. 
As $a$ is varied, the integrand
\bb
\sin^2 {a\pi\over k+2} [F(\cos {a\pi\over k+2})]
\ee
attains its maximum when
\bb
2\cot {a\pi\over k+2}=\sin {a\pi\over k+2}{F'\over F}.
\ee
At this maximum, $a$ satisfies
\bb
({a\pi\over k+2})^2\approx 2{F(1)\over F'(1)},
\ee
which is small because $F'$ contains $n_j$.
As ${a\pi\over k+2}$ is small, the integrand is approximated as
\bb
({a\pi\over k+2})^2[F(1-\frac12({a\pi\over k+2})^2)].
\ee
The width of the peak is estimated from the second derivative
\bb
\frac{2\pi^2}{(k+2)^2}F-\frac{5\pi^4a^2}{(k+2)^4}F'
+\frac{\pi^6a^4}{(k+2)^6}F'',
\ee
which, for large $n_j$, simplifies at the maximum to
\bb
-4\frac{\pi^2}{(k+2)^2}F.
\ee
Consequently the width $\sigma$ of the peak is given by
\bb
\sigma^2=
\frac{(k+2)^2}{\pi^2}\frac{F}{F'},
\ee
and the integral can be approximated by
\bb
\frac{2(k+2)}{\sqrt\pi}(\frac{F}{F'})^{3/2}F.
\ee
The number of states is
\bb
N=\frac{4}{\sqrt\pi}{(\sum_j n_j)!\over\prod_j n_j!}
(\sum_jn_j\frac{f_j'}{f_j})^{-3/2}\prod_j [f_j]^{n_j}.
\ee
To maximize this number, one sets 
\bb
\delta\log N=0
\ee
when the numbers $n_j$ of punctures with
spin $j$ are varied, 
subject to the constraint of fixed area
\bb
\sum_j\sqrt{j(j+1)}\delta n_j=0.
\ee
One obtains for large $n_j$
\bb
\log f_j+\log \sum_k n_k-\log n_j-\lambda\sqrt{j(j+1)}=0,
\ee
where $\lambda$ is a Lagrange multiplier. 
\bb
n_j=(\sum_k n_k)f_j e^{-\lambda\sqrt{j(j+1)}},
\ee
whence, for consistency,
\bb
\sum_{j=1/2}^{k/2} f_j e^{-\lambda\sqrt{j(j+1)}}=1.
\ee
$j$ goes from $\frac12$ to $\frac{k}{2}$ for
level $k$, infinite for large $k$ and
$f_j$ reduces to $2j+1$ as $\frac{a}{k+2}\to 0$
for large area.
This relation yields the same $\lambda$ as before. 

Taking the above distribution, one easily sees that
\bb
\log N=\frac{\lambda A}{8\pi\gamma\ell_P^2}
-\frac32\log A \quad {\rm for~}k\to\infty
\ee
as each $n_j$ goes like the area for large area. 

If however $k$ be fixed and the area made large,
the sum over the finite number of values of $a$
can be considered term by term.
For each $a$,
maximization of $N$ with respect to $n_j$ is as above,
but $(\frac{F}{F'})^{3/2}$ does not appear:
\bb
\log N=\frac{\lambda A}{8\pi\gamma\ell_P^2} \quad {\rm for~finite~} k.
\ee
Here $\lambda$ is determined with $f_j$ evaluated
for the $a$ under consideration and the sum restricted to $j\leq k/2$.
Now $\lambda$ depends on $k$ and also on $a$. 
In summation of $N$ over $a$,
highest $\lambda$ dominates and
has to be maximized over $a$, which determines
the relevant value(s) of $a$. No log corrections appear
because the $(\frac{F}{F'})^{3/2}$ factor, 
which appeared with $a$ taken to be continuous, 
does not appear in this case of finite $k$ and discrete $a$.

These calculations
are for most probable distribution, but the sum over all
distributions can be estimated. 
Correction factors
proportional to area from factorials and from integrations
approximating sums over $n_j$ cancel out
because there is only the area constraint.

\section*{Conclusion}
\begin{itemize}
\item The laws of black hole mechanics suggested that $S\propto A$.
\item Euclidean gravity indicated that $S=\frac{A}{4\ell_P^2}$.
\item Loop quantum gravity has indicated that 
\framebox{$S=\frac{A}{4\ell_P^2}-\frac12\ln A$}
for area eigenvalues $A$ close to the classical horizon area,
with a suitable choice of the Barbero-Immirzi parameter.
\item However, \framebox{$S=\frac{A}{4\ell_P^2}$}
for large $A$ and fixed classical area again
with a suitable choice of the Barbero-Immirzi parameter.
\end{itemize}

\end{document}